\documentstyle[twocolumn,aps,prl]{revtex}

\begin{document}

\draft
\flushbottom
\twocolumn[\hsize\textwidth\columnwidth\hsize\csname @twocolumnfalse\endcsname




\title{Chirality in Bare and Passivated Gold Nanoclusters}

\author{
   I.~L.~Garz\'on$^1$,
   J.~A.~Reyes-Nava$^1$,
   J.~I.~Rodr\'{\i}guez-Hern\'andez$^2$,
   I.~Sigal$^1$,
   M.~R.~Beltr\'an$^2$, and
   K.~Michaelian$^1$
   }
\address{
   $^1$Instituto de F\'{\i}sica,
       Universidad Nacional Aut\'onoma de M\'exico,
       Apartado Postal 20-364, M\'exico D.F., 01000 M\'exico \\
   $^2$Instituto de Investigaciones en Materiales,
       Universidad Nacional Aut\'onoma de M\'exico,
       Apartado Postal 70-360, M\'exico D.F., 01000 M\'exico \\
      }
\date{\today}

\maketitle

\begin{abstract}
Chiral structures have been found as the lowest-energy isomers
of bare (Au$_{28}$ and Au$_{55}$) and thiol-passivated 
(Au$_{28}$(SCH$_{3}$)$_{16}$ and Au$_{38}$(SCH$_{3}$)$_{24}$) 
gold nanoclusters. The degree of chirality existing in the 
chiral clusters was calculated using the Hausdorff chirality
measure. 
We found that the index of chirality is higher in the passivated 
clusters and decreases with the cluster size.
These results are consistent with the observed chiroptical 
activity recently reported for glutahione-passivated 
gold nanoclusters, and provide theoretical support for
the existence of chirality in these novel compounds.
\end{abstract}


]

Detailed knowledge of the lattice structure, shape, morphology,
surface structure, and bonding of bare
and passivated gold clusters
is fundamental to predict
and understand their electronic, optical, and other
physical and chemical properties.
This information is essential to optimizing their utilization
as novel nanocatalysts \cite{Cata}, and  
as building-blocks of new molecular nanostructured materials \cite{Whet1},
with potential applications in nanoelectronics \cite{Andres1}
and biological diagnostics \cite{Zan1}.
An effective theoretical approach to determine gold cluster structures
is to combine
genetic algorithms and many-body
potentials (to perform global structural optimizations), and first principles 
density functional theory (to confirm the energy ordering of the local 
minima). Using this procedure
we recently found many topologically interesting
\it disordered \rm gold nanoclusters
with energy near or below the lowest-energy ordered isomer
\cite{IG1,IG2,KM1,JS1}.
The structures of these clusters showed low spatial
symmetry or no symmetry at all, opening the possibility
of having distinct electronic and optical properties in such systems.
In other studies 
on passivated gold nanoclusters \cite{IG3,IG4},
we also found that
the effect of a methylthiol monolayer (24 \, SCH$_{3}$ molecules) 
on a truncated-octahedron (with fcc geometry) Au$_{38}$ cluster is strong
enough to produce a dramatic distortion on the gold cluster,
resulting in
a \it disordered \rm geometry for the most stable
Au$_{38}$(SCH$_{3}$)$_{24}$ passivated cluster.

Although the calculated structure factors
of the disordered gold clusters were in qualitative agreement with
the data obtained from X-ray powder 
diffraction on experimental samples \cite{IG2,KM1},
the direct confirmation of the existence of bare and
thiol-passivated gold nanoclusters with low or no spatial symmetry
had not been possible 
due to the lack of enough experimental resolution
for clusters in the size range of 1-2 nm \cite{Lan1,Whet2,Ugar1}.
Nevertheless, in a recent study using circular dichroism,
Shaaff and Whetten (SW) \cite{Whet3}
found a strong optical activity in the metal-based electronic
transitions (across the near-infrared, visible and near ultraviolet regions)
of size-separated glutathione-passivated gold clusters in the size
range of 20-40  Au atoms.
SW pointed out that the most plausible interpretation of these results
is that the structure of the
metal-cluster core of the gold-glutathione cluster compounds would be 
inherently chiral. Moreover, since the most abundant cluster in
the experimental samples corresponds to
the passivated cluster
Au$_{28}$(SG)$_{16}$, where SG denote the gluthatione
adsorbate, SW proposed a chiral structure, with $T$ point group,
for the Au$_{28}$ cluster \cite{Whet3}.

In earlier studies \cite{IG1,IG2,KM1,JS1,IG3,IG4}, we have reported
several structural, vibrational and electronic properties 
of the disordered gold nanoclusters. However,
although an initial quantification of the amount and type of 
local disorder present in amorphous-like structures of Au$_{55}$ 
was obtained \cite{IG5}, to our knowledge, no attempt to theoretically
investigate the existence of
chirality in gold nanoclusters has been done before.
In the light of the results obtained by SW regarding the intense chiroptical
activity of size-separated glutathione-passivated gold nanoclusters, it is
of interest to determine if the most-stable disordered structures
we have found for bare and passivated gold nanoclusters are chiral, 
and to quantify their degree of chirality. 
In this letter, we present a structural analysis 
that shows
the existence of chirality in these nanostructures.
This is done by implementing a method, based on 
the Hausdorff chirality measure 
\cite{Hauss1,Hauss2}, to quantify
the index of chirality of clusters.
Our results show that the metal cores of the passivated clusters
are more chiral than the bare clusters
and that the degree of chirality
decreases with the cluster size. 
These results provide
new insights on the effect of passivating agents on the bare metal
cluster as a mechanism to generate chiral nanostructures with
novel and perhaps unexpected properties.

Our initial step was to determine the lowest-energy structures of the
bare and passivated gold nanoclusters. To this end we first performed global
optimizations using the Gupta n-body potential and a genetic-symbiotic
algorithm for bare gold nanoclusters of different sizes
(Ref. \cite{KM1} provides details on this methodology) . Through this
procedure we obtained the distribution in energy of the most stable isomers
for each cluster size. From this distribution, we selected representative
isomers like those with the lowest-energy,
and those isomers that were considered
good candidates to be the lowest-energy minima based on the existence
of well known symmetric structures for certain cluster sizes
like the truncated octahedron, icosahedron, decahedron, etc.
The representative isomers were then
locally reoptimized through unconstrained relaxations
using forces calculated from density functional theory (DFT) in the
local-density (LDA) and generalized-gradient (GGA) approximations.
The DFT calculations were done using the  
{\sc Siesta} code \cite{Arta99}
with scalar-relativistic  norm-conserving pseudopotentials
\cite{tm2} and a double-$\zeta$ basis set of numerical
atomic orbitals \cite{Arta99} 
(see Refs. \cite{JS1,IG4} for additional details on these calculations).

Here, we report results  
for the Au$_{28}$ cluster since the lowest-energy structures
for the Au$_{n}$, $n$=38,55,75, clusters have been published elsewhere
\cite{IG1,IG2,KM1,JS1}. Our results for the Au$_{28}$ cluster show a
similar pattern as was found for the larger sizes: the distribution of
low-energy isomers shows a set of nearly degenerate in energy disordered
structures, where the lowest-energy one corresponds to a highly distorted
geometry. 
The $T$ structure proposed by SW \cite{Whet3} was obtained as a local
minimum of the potential energy surface with higher energy.
The top panel of Fig. 1 shows both cluster geometries and the distribution
of distances from the cluster center of mass. These structures
mainly differ in the presence of a 3-atom core
in the lowest-energy disordered isomer whereas the $T$ geometry contains
a tetrahedral core.
The DFT-LDA-GGA results confirm the higher stability of our
disordered structure as compared with the $T$ structure: 
we obtained
an energy difference of -0.885 eV (LDA) and -0.638 eV (GGA)
between both isomers.
A similar
trend, showing the higher stability of disordered isomers with
respect to ordered ones, has been obtained for larger cluster sizes: 
Au$_{n}$, $n$=38,55,75 \cite{IG1,JS1}.
The physical origin of the higher stability of amorphous-like 
structures in bare gold nanoclusters has been studied by analyzing the
metallic bonding and its effect on the local stress 
of the clusters. It 
results that the amorphization
is a mechanism for strain relief that lowers the
cluster energy, being more prominent in the case of
gold due to the short-range of the n-body metallic
interaction \cite{KM1,JS1,JS2}.

The lowest-energy structures of the thiol-passivated 
Au$_{28}$(SCH$_{3}$)$_{16}$
and Au$_{38}$(SCH$_{3}$)$_{24}$ clusters were obtained
by performing local relaxations, 
using the forces calculated from the DFT-LDA-GGA first principles method,
and starting from different cluster-monolayer configurations.
These include the lowest-energy ordered and disordered
bare Au$_{28}$ and Au$_{38}$ cluster geometries obtained
by the procedure described above, with the methylthiols molecules
placed on different adsorption sites (top, bridge, hollow),
as well as on random positions over the metal cluster surface.
Figure 2 (bottom panel) shows a highly distorted passivated cluster
that corresponds to the lowest-energy structure of
the Au$_{28}$(SCH$_{3}$)$_{16}$ cluster, obtained by the relaxation
of the disordered bare Au$_{28}$ cluster with the thiols placed
close to 3-atom hollow sites. 
From the distribution of radial distances to the cluster center of mass
shown in Fig. 2 (bottom panel), the degree of distortion and
the ill-defined nature of the gold-thiol interface is evident.
A similar strongly
distorted geometry was found for the lowest-energy
structure of the larger passivated Au$_{38}$(SCH$_{3}$)$_{24}$
cluster \cite{IG3,IG4}. 
The main driving force producing
the cluster distortion is the strong gold-sulphur interaction.
An analysis of the metal-ligand interaction in the gold passivated clusters
will be published elsewhere \cite{IG6}. Meanwhile, recent calculations
on smaller Au-thiolate compounds have confirmed 
that the strong covalent (directional)
gold-sulphur interaction is  mainly-responsible for the
metal cluster distortion \cite{Kru}.

The principal objective of this work is to determine if the disordered
lowest-energy structures of the bare Au$_{n}$, $n$= 28,38,55,75,
and passivated Au$_{28}$(SCH$_{3}$)$_{16}$ and 
Au$_{38}$(SCH$_{3}$)$_{24}$ clusters are chiral, and to
quantify the degree of chirality existing in them.
This information is relevant for a proper interpretation of the
circular dichroism measurements performed by SW 
on glutathione-passivated gold clusters \cite{Whet3}.
Since chirality is a geometrical property of the system, independent
of its chemical and physical manifestations, it is possible to
quantify chirality without reference to experimental measurements,
but using the inherent structural symmetry of the clusters.
Although in recent years several approaches have been developed
to measure chirality \cite{Weinberg,Harris},
the Hausdorff chirality measure,
proposed by Buda and Mislow (BM) \cite{Hauss1}, 
has emerged as the general method of choice for the quantification
of chirality \cite{Hauss2}. 
Within this approach, 
the degree of chirality is found
by calculating the maximum overlap between the actual molecular structure
and its mirror image, using the Hausdorff distance \cite{HD} between their
sets of atomic coordinates. By rotating and translating one
structure with respect to the other, the optimal overlap
can be calculated. 
This Hausdorff chirality
measure (HCM) is a continuous and similarity invariant function of the
molecular shape, and is zero only if the molecule is achiral.
The advantage of this approach is that its
numerical implementation for large clusters containing
\it n \rm atoms in a three-dimensional space is straightforward, 
as was discussed by BM \cite{Hauss1,Hauss2}.
Our numerical procedure was tested by calculating the degree of chirality
of the three tetrahedra shapes considered by BM, obtaining the same values
as reported by these authors \cite{Hauss1}.

We calculate the HCM for the lowest-energy structures of the bare and
thiol-passivated gold nanoclusters using their relaxed
Cartesian coordinates measured with respect to the cluster center of mass. 
Through an inversion operation, the coordinates
of the mirror image clusters were obtained. The HCM was obtained
by calculating the maximum overlap between a given cluster and its mirror
image. This corresponds to the minimum value of the Hausdorff
distance between the sets of atomic coordinates of both structures.
In order to obtain the minimum Hausdorff distance, the mirror
cluster was translated and rotated around the original cluster in the
three-dimensional space generating different configurations. 
For each of them, the Hausdorff distance with respect to the original
cluster was calculated.
The minimum of these values, normalized
by the largest interatomic distance in the cluster, corresponds to the
HCM. To check the reliablity of this methodology, the
minimization of the Hausdorff distance was performed using the
Broyden-Fletcher-Goldfarb-Shano method as suggested in
Refs.\cite{Hauss1,Hauss2}, a conjugate gradient optimization,
and a direct evaluation of the Hausdorff distance
taking variations of  $\pi/360$ for
each rotational angle. The HCM values agree for the three methods
up to the third significant digit.

Our results show that the HCM index of chirality
for Au$_{38}$ and Au$_{75}$ is zero. These values are expected since
the lowest-energy structures corresponding to these sizes have
one plane of symmetry and therefore are achiral. In fact, 
these results constitute an additional
test of our numerical implementation for calculating HCM in large clusters.
The HCM index of chirality for the lowest-energy structure and the
$T$ isomer proposed by SW for Au$_{28}$ are displayed in Table I, together
with the value obtained for the lowest-energy structure
of Au$_{55}$ (see Fig. 1 in Ref. \cite{IG5}). 
These results show that the most stable structure
of Au$_{28}$, obtained in this work, is less chiral than the one
proposed by SW, and that the index of chirality varies slightly with the
cluster size from 28 to 55 atoms.
For the calculation of the HCM of the thiol-passivated gold nanoclusters,
we only consider the coordinates of the Au atoms since we are interested
in testing the suggestion of SW who proposed that the metal cluster core
would be inherently chiral \cite{Whet3}. The HCM values for the
metal core of the passivated 
Au$_{28}$(SCH$_{3}$)$_{16}$ and
Au$_{38}$(SCH$_{3}$)$_{24}$ clusters are also shown in Table I.
These results show that both metal cluster cores are more chiral
than the bare Au$_{28}$ and Au$_{38}$ clusters, and
that the index of chirality for the passivated clusters decreases 
with the cluster size.
It is noteworthy that the HCM values obtained for the bare
and passivated chiral Au clusters are smaller than the values of
the chiral tetrahedral shapes (HCM = 0.221, 0.252, 0.255 for $D_2$, $C_2$,
and $C_1$ tetrahedra, respectively) obtained  by BM \cite{Hauss1}.
In order to compare the HCM of Au clusters with a physical system, 
we calculated the index of chirality
of the chiral $D_2$-C$_{76}$ and $D_2$-C$_{84}$, and the achiral
$I_h$-C$_{60}$ fullerenes, obtaining the values
0.109, 0.102, and 0.000, respectively.
These results support the reliability of our methodology
since there is experimental evidence on the 
existence of chirality in
the $D_2$-C$_{76}$ fullerene \cite{Fuller}.
This comparison also provides a useful reference to classify 
cluster and molecules
in terms of their indexes of chirality, and indicates
that the bare Au$_{28}$ and Au$_{55}$ clusters are as chiral as the 
$D_2$-C$_{76}$, but the passivated Au$_{28}$(SCH$_{3}$)$_{16}$ and
Au$_{38}$(SCH$_{3}$)$_{24}$ clusters are slightly more chiral
than such fullerene.

The above results on the existence of chirality
in thiol-passivated Au nanoclusters are consistent with 
the chiroptical activity
measured on size-separated glutathione-passivated gold
nanoclusters in the size-range 20-40 Au atoms since
it was shown that the optical properties are insensitive to the
tail of the adsorbate thiol \cite{Whet3}.
Also, in the experimental samples it was found that the optical activity is
weaker as the cluster size increases, and it is lost for larger
clusters and for samples with mixed cluster sizes \cite{Whet3}. 
It remains to be verified if indeed there is a correlation between the
value of the index of chirality and the strength of the intensity
or other spectral features of the optical activity signal in the
chiral samples.
Another interesting
prediction from the present calculations, to be confirmed
experimentally, is related with the smaller index of chirality
existing on bare clusters as compared to those 
for the metal-core of passivated clusters. 
Our results imply that the effect of the passivating monolayer
is strong enough to distort a bare cluster geometry, producing chiral
metal-cores that give rise to the intense chiroptical activity.
This effect could change an achiral cluster into a chiral one,
like in the Au$_{38}$ case, or increase the index of chirality
in an already chiral structure, like the Au$_{28}$ cluster.

The main contribution of this theoretical
work has been to show the existence of, and 
quantify chirality in the lowest-energy
structures of gold nanoclusters using the Hausdorff
chirality measure. These results provide useful insights for a proper
interpretation of the experimental results that report optical activity
on passivated gold nanoclusters \cite{Whet3}. 
Our results also establish that passivating with thiol monolayers
provides a mechanism to
induce or increase chirality in bare Au clusters in the size-range
of 20-40 atoms. This new information is relevant for the growth
and sample preparation of different chiral metal nanoclusters.
Theoretical precedents on 
metallic chiral nanostructures would include the
Ni$_{28}$ cluster \cite{Pristo}, and the Al and Pb helical
nanowires \cite{Tosa}, however the present results and those
obtained by SW provide both theoretical and experimental evidence
on the existence of chirality in gold nanoclusters.
Electronic, optical, thermal, and other physical and chemical properties
of chiral gold nanoclusters are currently being calculated and
will be reported in forthcoming publications.
At present, it is expected that
novel and interesting properties emerge from the chiral character
of metal clusters that could be useful for new
nanotechnological applications.

\begin{acknowledgements}
This work was supported by 
Mexico's CONACYT grant 28822-E, and the
DGSCA-UNAM Supercomputing Center.
\end{acknowledgements}

\bigskip
\bigskip
\bigskip
\bigskip
\bigskip
\begin{table}
\caption{Index of chirality of the lowest-energy
structures of bare (first column) and thiol passivated (third column)
gold nanoclusters.The second column shows that all the ordered isomers
are achiral except the $T$ isomer of Au$_{28}$.
The index of chirality was calculated using the Hausdorff distance
between the sets of atomic coordinates of the cluster and its
mirror image. The cluster
geometry was obtained through structural optimizations using a many-body
potential, genetic algorithms and density functional theory.}
\begin{tabular}{c|c|c}
\multicolumn{2}{c}{Bare} & Passivated \\
\hline
Au$_{28}$ (disordered)& Au$_{28}$ ($T$) & 
Au$_{28}$(SCH$_{3}$)$_{16}$ (disordered) \\
0.106 & 0.129 & 0.160 \\
\hline
Au$_{38}$ ($C_s$) & Au$_{38}$ ($O_h$) & 
Au$_{38}$(SCH$_{3}$)$_{24}$ (disordered)\\
0.000 & 0.000 & 0.121 \\
\hline
Au$_{55}$ (disordered) & Au$_{55}$ ($I_{h}$) &   \\
0.117 & 0.000 & \\
\hline
Au$_{75}$ ($C_s$) & Au$_{75}$ ($D_{5h}$) &  \\
0.000 & 0.000 &  \\
\hline
\end{tabular}
\end{table}
\bigskip
{\bf FIGURE CAPTIONS}
\begin{description}

\item[FIG. 1.]
Top panel: Distances of the gold  atoms
from the center of mass and cluster geometry for the lowest-energy
disordered (closed diamonds and inset (a)) and the $T$
(open circles and inset (b)) isomers of the bare Au$_{28}$ cluster.
Bottom panel: Distances of the gold (open diamonds)
and sulphur (stars) atoms
from the center of mass for the lowest-energy structure of
the thiol-passivated Au$_{28}$(SCH$_{3}$)$_{16}$ cluster.
The closed diamonds denote the same gold atom distances as in the
top panel. They are included to show at the same
scale, the degree of distortion and expansion
of the gold metal cluster core upon passivation.
The insets show the geometries of the metal core and the passivated cluster.
Sulphur atoms are depicted as darker spheres. 
\end{description}


\begin{references}
\bibitem{HD} The Hausdorff distance $h(Q,Q')$  between the
sets $Q$ and $Q'$ can be defined as
the smallest number $\delta = h(Q,Q')$ that has
the following properties: (a) a spherical ball of radius 
$\delta$ centered at any point of Q contains at least one point
of $Q'$, and (b) a spherical ball of radius $\delta$ centered at
any point of $Q'$ contains at least one point of $Q$. This
definition was given in Refs. \cite{Hauss1,Hauss2}.
\end{references}
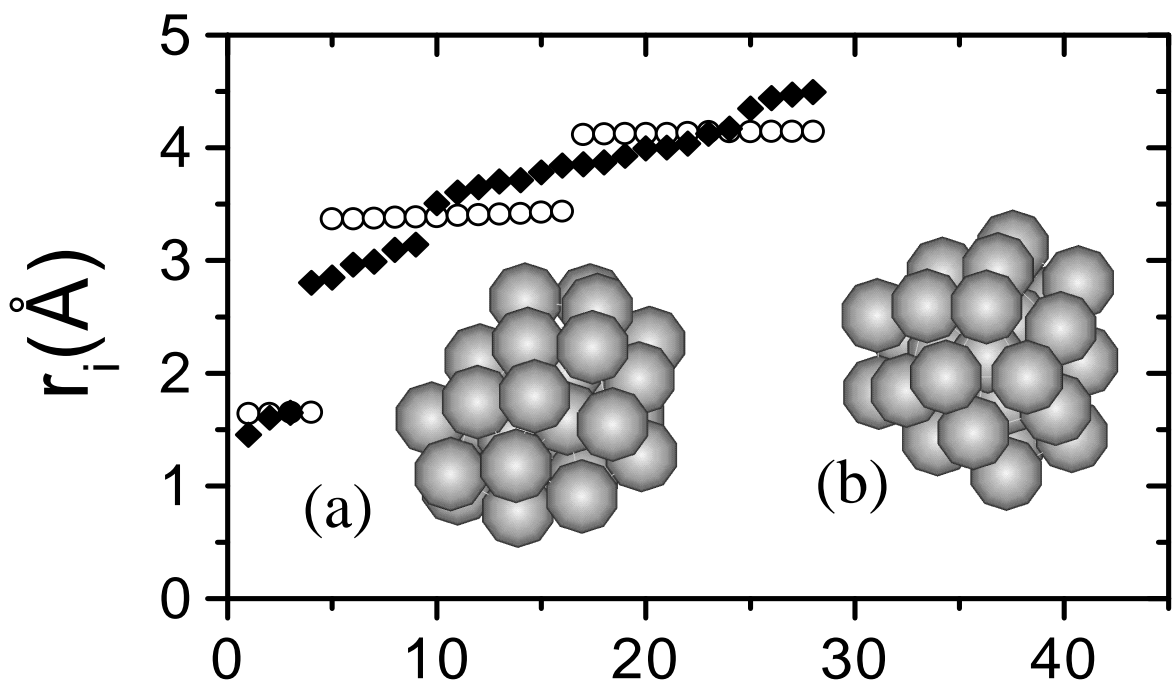
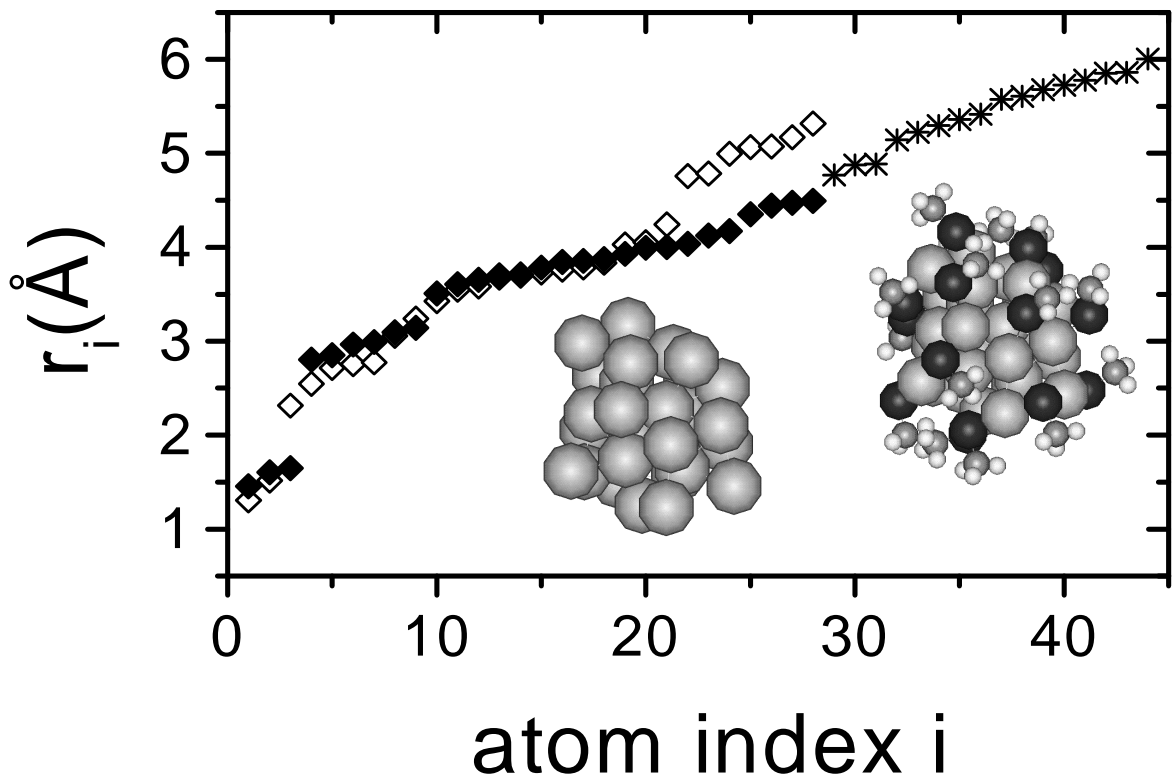
\end{document}